\documentclass[12pt]{article}
\usepackage{pdproc,epsfig}

\newcommand{\eg}{{e.g.,\hskip 3pt}}

\def\4he{$^4$He}
\def\3he{$^3$He}
\def\7li{$^7$Li}
\def\6li{$^6$Li}
\def\Yp{Y$_{\rm P}$~}

\def\hii{H\thinspace{$\scriptstyle{\rm II}$}~}

\def\Nnu{N$_{\nu}$~}

\newcommand{\epm}{\ensuremath{e^{\pm}\;}}
\newcommand{\Deln}{\ensuremath{\Delta N_\nu\;}}
\def\Nnu{N$_{\nu}$~}

\newcommand{\be}{\begin{equation}}
\newcommand{\ee}{\end{equation}}
\newcommand\la{\lower0.6ex\vbox{\hbox{\ensuremath{\buildrel{\textstyle<}\over{\sim}\ }}}}
\newcommand\ga{\lower0.6ex\vbox{\hbox{\ensuremath{\buildrel{\textstyle>}\over{\sim}\ }}}}

\begin{document}

\renewcommand{\thefootnote}{\alph{footnote}}
  
\title{NEUTRINOS AND BBN (AND THE CMB)}

\author{ GARY STEIGMAN}

\address{ Physics Department, The Ohio State University\\  191 West Woodruff
Avenue, Columbus, OH 43210, USA\\
 {\rm E-mail: steigman@mps.ohio-state.edu}}

\abstract{During Big Bang Nucleosynthesis (BBN), in the first $\sim$20 
minutes of the evolution of the Universe, the light nuclides, D, \3he, 
\4he, and \7li were synthesized in astrophysically interesting abundances.  
The Cosmic Microwave Background Radiation (CMB) observed at present was 
last scattered some $\sim$400 thousand years later.  BBN and the CMB 
(supplemented by more recent, $\sim$10 Gyr, Large Scale Structure data), 
provide complementary probes of the early evolution of the Universe and 
enable constraints on the high temperature/energy physical processes in 
it.  In this overview the predictions and observations of two physical 
quantities, the baryon density parameter and the expansion rate parameter, 
are compared to see if there is agreement between theory and observation 
at these two widely separated epochs.  After answering this question in 
the affirmative, the consequences of this concordance for physics beyond 
the standard models of particle physics and cosmology is discussed.}
   
\normalsize\baselineskip=15pt

\section{Introduction}

Big Bang Nucleosynthesis (BBN) probes the evolution of the Universe
during its first few minutes, providing a glimpse into its earliest
epochs.  The photons in the presently observed Cosmic Microwave 
Background Radiation (CMB) were last scattered at``recombination", 
when the plasma of ions and electrons became predominantly neutral, 
some 400 thousand years later.  Comparison of observations relating 
to BBN (the primordial abundances of the light elements) and the CMB 
(the temperature anisotropy spectrum) often provide unique tests of 
the consistency of the standard models of particle physics and cosmology, 
as well as enable the possibility of constraining alternate models of 
cosmology and of physics beyond the standard model of particle physics.  
In this article, based on my invited talk at the NO-VE IV International 
Workshop on: ``Neutrino Oscillations in Venice", the complementarity 
between BBN and the CMB (supplemented by the Large Scale Structure (LSS) 
data needed to break some degeneracies) is explored, concentrating on 
the universal baryon asymmetry and the expansion rate of the early, 
radiation-dominated Universe.  In particular, do observations of the 
baryon density parameter and of the expansion rate parameter (both 
to be defined in more detail below) agree at $\sim$20 minutes and 
$\sim$400 thousand years and, if so, what does this tell us about 
non-standard physics and/or cosmology in the time interval between 
these two, so widely separated, epochs?

After introducing and defining the cosmological parameters in \S1.1 
and \S1.2, their influence on the predictions of BBN (\S2) and the CMB 
observations (\S3) is discussed and compared to the standard model 
predictions and, with the observational data (\S4).  After establishing 
the concordance of the standard models of particle physics and cosmology, 
some general constraints on possible new physics in the interval 
between BBN and recombination are presented in \S5.  The results are
summarized in \S6.  In my recent review of BBN\cite{steigman07} these 
issues are discussed in more detail (including a more extensive list 
of references).  My talk at NO-VE IV and this article are drawn from 
the recently published paper with V. Simha\cite{vs1}.

\subsection{The Baryon Density Parameter: $\eta_{\rm B}$}

During the much earlier evolution of the Universe than is considered
here, a universal matter-antimatter (baryon-antibaryon) asymmetry was 
established by particle physics processes yet to be uniquely determined.  
Thereafter, in the evolution of the Universe up to the present, the number 
of baryons in a comoving volume is preserved, although the number density 
of baryons decreases as the Universe expands.  During the same early 
evolution of the Universe, very rapid electromagnetic processes establish 
and maintain a Bose-Einstein (black body) distribution for the cosmic 
background photons.  This guarantees that when the photons are in 
equilibrium or, when they are entirely decoupled, the number of CMB 
photons in a comoving volume is preserved (except for the new photons 
added when various particle-antiparticle pairs annihilate or, when unstable 
particles decay).  In the standard models of cosmology and particle 
physics the numbers of baryons and of CMB photons in a comoving volume 
have been unchanged since \epm anniliation.  The ratio of the numbers 
of baryons and CMB photons in a comoving volume in the post-\epm 
annihilation Universe provides us with a dimensionless, time-invariant 
parameter -- in the context of the {\bf standard} models of particle 
physics (baryon conservation) and cosmology (entropy conservation).
\be
\eta_{\rm B} \equiv n_{\rm B}/n_{\gamma} \equiv 10^{-10}\eta_{10}.
\ee
$\eta_{10}$ is related to the baryon mass density parameter $\Omega_{\rm B}$, 
the ratio of the baryon mass density to the critical mass density, 
and the reduced, present value of the Hubble parameter $h$ (the 
Hubble ``constant": $H_{0} \equiv 100h$~km~s$^{-1}$~Mpc$^{-1}$) 
by\cite{steigman06}, 
\be
\eta_{10} = 273.9~\Omega_{\rm B}h^2.
\ee
$\eta_{10}$ ``measured" at BBN ($\sim$20 minutes) and inferred from the 
CMB ($\sim$400 thousand years later) should agree, enabling constraints 
to be placed on non-standard entropy production and/or baryon number 
violation.  

\subsection{The Universal Expansion Rate Parameter: $S$}

In the standard model of cosmology, the expansion rate is described 
by the Hubble parameter ($H$) which, in the early Universe, is related 
to the total energy density through the Friedman equation by,  
\be
H^{2} = 8\pi G\rho_{\rm TOT}/3.
\label{hubble}
\ee
During its early evolution, the Universe is ``radiation dominated"
($\rho_{\rm TOT} \rightarrow \rho_{\rm R}$); that is, the total energy 
density is dominated by massless particles (\eg photons) or relativistic, 
massive particles ($m~\la T$).  Prior to \epm annihilation, for temperatures 
$T~\la$~few MeV, the standard-model relativistic particles present are 
photons, \epm pairs, and three flavors of left-handed neutrinos along 
with their right-handed antineutrinos.  Counting degrees of freedom 
and accounting for the relativistic bosons and fermions,
\be
\rho_{\rm R} = \rho_{\gamma} + \rho_{e} + 3\rho_{\nu} = (1 + 7/8(2 + 3))\rho_{\gamma} 
= 43\rho_{\gamma}/8.
\ee

As may be seen from eq.~\ref{hubble}, any modification to the Friedman 
equation will change the expansion rate ($H \rightarrow H' \equiv SH$).  
For example, a non-standard, early-Universe expansion rate may be due 
to the presence of a non-standard energy density of relativistic particles 
($\rho_{\rm R}' \neq \rho_{\rm R}$), or by a non-standard value of the 
gravitational constant $G' \neq G$), or by some form of new physics beyond 
the standard model which modifies the Friedman equation.  A non-standard 
particle content may be conveniently parameterized by the equivalent 
number of ``additional" neutrinos defined, prior to \epm annihilation, 
by $\Delta$N$_{\nu} \equiv$ N$_{\nu} - 3$, so that $\rho_{\rm R}' \equiv 
\rho_{\rm R} + \Delta$N$_{\nu}\rho_{\nu}$. \Deln and the expansion rate
parameter $S$ are related by\cite{steigman07},
\be
S^{2} \equiv \left({H' \over H}\right)^{2} = {\rho_{\rm R}' \over \rho_{\rm R}} = 
{G' \over G} \equiv 1 + {7\Delta{\rm N}_{\nu} \over 43} = 1 + 0.163\Delta{\rm N}_{\nu}.
\ee 
However, {\bf any} change in $H$ ($S \neq 1$) which is traceable to a term 
evolving like the radiation density (as the inverse fourth power of the 
scale factor) can equally well be parameterized by $\Delta$N$_{\nu}$.  In 
this sense, \Deln $\neq 0$ should be thought of as a proxy for any non-standard, 
early-Universe expansion rate ($S \neq 1$) and need have {\bf nothing at 
all} to do with ``extra" (or fewer!) neutrinos.

After \epm annihilation the only relativistic particles present during
the radiation dominated epoch are the photons (which redshift to become 
the presently observed CMB) and the neutrinos, which decoupled prior
to \epm annihilation.  In the approximation that the neutrinos were
{\bf fully} decoupled at \epm annihilation, the post-annihilation photons 
are hotter than the neutrinos by a factor of $T_{\gamma}/T_{\nu} 
= (11/4)^{1/3}$ and\cite{steigman07}
\be
S^{2} = \rho'_{\rm R}/\rho_{\rm R} \rightarrow 1 + 0.135 \Delta {\rm N}_{\nu},
\ee
where, in the post-BBN, pre-recombination Universe, $\Delta$N$_{\nu} 
\equiv$~N$_{\nu} - 3$.

Since the standard-model neutrinos were {\bf not} fully decoupled at \epm 
annihilation, they share some of the energy/entropy released during 
\epm annihilation\cite{dicus}.  As a result, they are warmer than is 
predicted by the fully decoupled approximation, increasing the ratio 
of the post-\epm annihilation radiation density to the photon energy 
density.  In the standard model, this additional contribution to the 
total energy density can be accounted for by replacing \Nnu = 3 with 
\Nnu = 3.046\cite{mangano}.  Any post-BBN deviations from the standard 
model that can be treated as equivalent to contributions from {\bf 
fully decoupled} neutrinos can thus be accounted for by\cite{steigman07},
\be
S^{2} = \rho'_{\rm R}/\rho_{\rm R} \rightarrow 1 + 0.134\Delta {\rm N}_{\nu},
\ee
where in the post-\epm annihilation Universe relevant for comparison 
with the CMB and LSS, $\Delta$N$_{\nu} \equiv$~N$'_{\nu} - 3.046$.  
Note that with these definitions, the standard model corresponds 
to \Deln = 0 in {\bf both} the pre- and post-\epm annihilation Universe.  

\section{Overview Of Primordial Nucleosynthesis}
\label{BBN}

The early, hot, dense Universe evolves through a brief epoch when it 
is a cosmic nuclear reactor.  Since for ``our" Universe, the entropy 
per baryon (nucleon) is very large, $\sim$10$^{9}$, the synthesis of 
the elements is delayed well beyond the time when the average thermal 
energy ($\propto T$, the temperature of the radiation and of those 
particles in equilibrium with it) drops below the binding energy of 
the lightest nuclei, in particular, deuterium.  As a result, even 
though nuclear reactions such as $n + p \rightarrow {\rm D} + \gamma$ 
begin very early, due to the large $\gamma$-ray background provided 
by the ``blue-shifted CMB", the ``back reaction" ${\rm D} + \gamma 
\rightarrow n + p$, keeps the deuterium abundance very small, 
inhibiting the formation of any of the more complex nuclei until 
later, at $t \sim 3$~minutes, when $T \approx 0.08$~MeV.  At this 
time the number density of those photons with sufficient energy to 
photodissociate deuterium is comparable to the number density of 
baryons and only now can the various two-body nuclear reactions 
begin to build more complex nuclei such as \3he ($^{3}$H), \4he, 
and \7li ($^{7}$Be).  The absence of a stable nuclide at mass-5 
ensures that the primordial abundance of \7li is much smaller than 
that of the other light nuclides and the similar gap at mass-8 
guarantees negligible primordial abundances for any heavier nuclei.  
Less than $\sim$20 minutes later, when the temperature in the expanding, 
cooling Universe has dropped to $T \approx 0.03$~MeV, the coulomb 
barriers in collisions between charged nuclei, combined with the 
absence of free neutrons (most of which have, by now, been incorporated 
into \4he, the most tightly bound of the light nuclei), nuclear 
reactions end and the primordial abundances of the light nuclei 
are frozen out.  

For nucleosynthesis in the standard cosmology (SBBN), there is only 
one adjustable parameter, the baryon density parameter $\eta_{\rm B}$ 
(or, $\eta_{10}$).  Observations which lead to a determination of the 
primordial abundance of any one of the light nuclides can determine 
$\eta_{10}$, which then may be compared with its value inferred from 
the CMB.  The internal consistency of SBBN can be checked by comparing 
the abundances of the other nuclides, predicted using this value of 
$\eta_{10}$, with the observationally-inferred abundances.  However, 
in constrast to the other light nuclides, the BBN-predicted primordial 
abundance of \4he is very insensitive to the baryon density parameter.  
Rather, the \4he mass fraction, Y$_{\rm P}$, depends on the neutron-to-proton 
ratio at BBN since virtually all neutrons available at that time are 
incorporated into \4he.  In turn, the n/p ratio depends on the competition 
between the charged-current weak interaction rate (normalized, for 
example, by the accurately known neutron lifetime) and the universal 
expansion rate, parameterized by $S$ (or N$_{\nu}$).  Therefore, while 
D, \3he, and \7li are potential baryometers, \4he provides a potential 
chronometer.

\subsection{Deuterium}

Of the light nuclides produced in astrophysically interesting abundances,
deuterium is the baryometer of choice.  As the Universe evolves and gas 
is cycled through succeeding generations of stars, deuterium is only 
destroyed, so that its post-BBN evolution is simple and monotonic.  The 
abundance of deuterium observed anywhere, at any time in the evolution of 
the Universe, is never greater than its BBN abundance.  If D is observed 
in systems of low metallicity and/or or at high redshift, where very little 
gas has been cycled through stars, its observationally inferred abundance 
should approach its primordial value.  Furthermore, the BBN-predicted D
abundance is sensitive to the baryon density parameter ($y_{\rm DP} \equiv 
10^{5}$(D/H)$_{\rm P} \propto \eta_{\rm B}^{-1.6}$), so that a $\sim$10\% 
determination of $y_{\rm DP}$, would lead to a $\sim$6\% estimate of 
$\eta_{10}$.

\begin{figure}
\centerline{
\epsfxsize=3.25truein
\epsfbox[45 482 335 725]{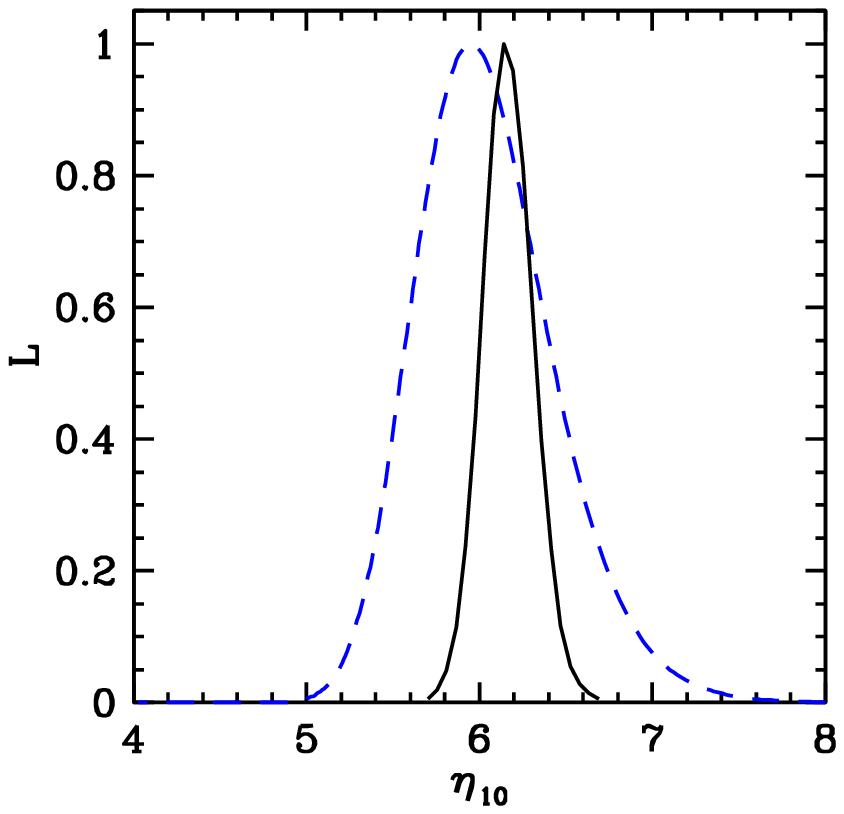}
}
\caption{}
The likelihood distributions of the baryon density parameter, $\eta_{10}$. 
The dashed (blue) curve shows the distribution inferred from SBBN (N$_{\nu} 
= 3$) and the adopted primordial abundance of deuterium.  The solid (black) 
curve is the distribution of $\eta_{10}$ for N$_{\nu} = 3$ inferred from the 
CMB using the WMAP 5-year and small scale CMB data, the LSS matter power 
spectrum and SNIa data, and the HST Key Project value of $H_{0}$.
\label{fig:eta10sbbn}
\end{figure}

Finding and observing suitable targets to determine $y_{\rm DP}$ is
telescope-intensive and subject to systematic errors.  As a result, at 
present, observations and reliable determinations of D/H are available 
for only 7, high-redshift, low-metallicity, QSO Absorption Line Systems, 
in which neutral D and H are observed in absorption against background
light sources (QSOs).  See Pettini et al. 2008\cite{pettini} for the most 
recent results and references to earlier work.  Unfortunately, given the 
errors quoted for the individual D/H determinations, the dispersion among 
their central values is excessive (\eg the reduced $\chi^{2}$ is~$\ga 3$), 
suggesting that the errors may have been underestimated and/or systematic 
effects contaminate one or more of the determinations.  Until this is 
resolved with more and/or better data, I adopt for $y_{\rm DP}$ the 
weighted mean of the individual D/H determinations, but inflate the 
errors to reflect the large dispersion (by multiplying them by the 
square root of the reduced $\chi^{2}$).  The result is that $y_{\rm DP} 
= 2.70^{+0.22}_{-0.20}$.  For SBBN (\Nnu = 3), this implies that 
$\eta_{10} = 5.96^{+0.30}_{-0.33}$, a $\sim$5\% determination of 
$\eta_{\rm B}$(BBN).  The SBBN D-inferred likelihood distribution of 
$\eta_{10}$ is shown as the dashed curve in Figure~\ref{fig:eta10sbbn}.
Before comparing this SBBN-inferred value of the baryon density 
parameter, measured at $\sim$20 minutes, with the value inferred
from the CMB at $\sim$400 thousand years, the self-consistency of 
SBBN is investigated considering the relic abundances of \3he, 
\4he and \7li.

\subsection{Helium-3}

For the D-inferred value of of $\eta_{10}$, the SBBN-predicted relic
abundance of \3he is $y_{3\rm P} \equiv 10^{5}(^{3}{\rm He}/{\rm H})_{\rm P} 
= 1.1\pm0.2$.  The post-BBN evolution of \3he is much more complicated 
-- and model dependent -- than that of deuterium (or, of \4he; see below).  
When gas is cycled through stars D is first burned to \3he.  In the 
cooler, outer layers of stars the prestellar D plus \3he is preserved.  
However, in the hotter stellar interiors \3he is burned away.  In general, 
the more abundant, cooler, lower mass stars are net producers of \3he, 
but it is unclear how much of this stellar-synthesized \3he is actually
returned to the interstellar medium (ISM).  For further details and 
references see my review article\cite{steigman07}.  The bottom line is 
that it is very difficult to use observations of \3he in chemically-evolved 
regions, such as the ISM of our Galaxy, to infer the relic abundance of \3he.  
Nonetheless, Bania, Rood \& Balser\cite{brb} adopt for an {\it upper} limit to 
the \3he primordial abundance, the abundance inferred from observations of 
the most distant (from the Galactic center), most metal poor (least chemically 
evolved?), Galactic \hii region, $y_{3\rm P}~\la 1.1\pm0.2$.  This is in 
excellent agreement (within the model-dependent uncertainties) with the 
SBBN prediction.

\subsection{Helium-4}

The SBBN-predicted \4he primordial mass fraction for the D-inferred 
value of $\eta_{10}$ is \Yp $= 0.2484 \pm 0.0007$.  As gas is cycled
through stars the post-BBN evolution of \4he is monotonic -- hydrogen
is burned to helium, increasing Y (along with the abundances of the 
heavier elements such as, \eg oxygen).  For observations of gas anywhere, 
at any time in the post-BBN evolution of the Universe, Y$~\ga$Y$_{\rm P}$.  
Since the BBN-predicted \4he relic mass fraction is very insensitive 
to the baryon density parameter, \4he is a poor baryometer.

In my opinion the observationally-determined value of \Yp (and its error) 
is currently unresolved.  The most recent analyses\cite{OS,FK,PLP}, using
selected subsets of the available data, fail to find evidence for the 
expected correlation between the helium and oxygen abundances, calling 
into question the model-dependent extrapolations to zero metallicity 
they employ to infer the primordial abundance of helium.  Here, the 
value suggested by me in my review\cite{steigman07} is chosen for the 
subsequent discussion, 
\be
{\rm Y}_{\rm P} = 0.240 \pm 0.006.
\ee
The adopted error is an attempt to account for the systematic, as well 
as the statistical, uncertainties.  While the central value of \Yp adopted 
here is low, it is only slightly more than 1$\sigma$ below the SBBN-predicted 
central value; within $\sim$1.4$\sigma$, the observations and predictions
agree.   

The recent data and analyses\cite{OS,FK,PLP} are, however, in agreement on 
a weighted mean of the post-BBN abundance, which can be used to provide 
an {\bf upper bound} to Y$_{\rm P}$.  To this end an alternate constraint 
on (upper bound to) Y$_{\rm P} < 0.251 \pm 0.002$ will also be used.  
At $\sim$2$\sigma$, this suggests that Y$_{\rm P} < 0.255$.  This upper 
bound to \Yp is entirely consistent with the SBBN-predicted abundance.

\subsection{Lithium-7}

For the D-inferred value of the baryon density parameter, the 
SBBN-predicted primordial abundance of lithium is [Li]$_{\rm P} 
\equiv 12 + $log(Li/H)$_{\rm P} = 2.63^{+0.06}_{-0.07}$.  Although 
the weakly-bound lithium nucleus is easily destroyed in the hot 
interiors of stars, theoretical expectations, supported by observational 
data, suggest that the overall post-BBN trend has been for lithium 
to increase its abundance in the Galaxy with time, and with increasing 
metallicity.  The key data for probing the BBN \7li yield are from 
observations of the surface abundances of the oldest, most metal-poor 
stars in the Galaxy.  These lithium abundances are expected to form 
a plateau, the so-called ``Spite plateau", in a plot of Li/H (or, of 
[Li]) versus metallicity.  The surprise from recent observational 
data\cite{ryan,aspl} is the apparent absence of evidence for a lithium 
plateau; the lithium abundance continues to decrease with decreasing 
metallicity.  It is difficult to know how to use these data to infer 
the primordial abundance of \7li.  The lowest metallicity data suggest 
that [Li]$_{\rm P}~\la 2.1\pm0.1$.  There is clearly tension between 
this estimate (upper bound?) and the SBBN-predicted relic abundance 
which differ by a factor of three or more.  The resolution of this 
conflict may be found in the stellar astrophysics: in the $\sim$13 Gyr 
lifetimes of the oldest stars in the Galaxy where lithium is observed, 
the surface abundances of lithium (and other elements) may have been 
transformed by mixing surface material with that from the hotter interior 
where lithium has been destroyed.  In support of this possibility, 
recent observations and stellar modeling by Korn {\it et al.}\cite{korn} 
are of interest.  They observe stars in a Globular Cluster which are 
of the same age and were born with the same metallicity.  Comparing 
their observations of lithium and of heavier elements to models of 
stellar diffusion, they find evidence that both lithium and the 
heavier elements may have settled out of the atmospheres of these 
stars.  Applying their stellar models to the data for this set of 
stars, they infer for the unevolved lithium abundance, [Li] = 2.54 $\pm 
0.10$, in excellent agreement with the SBBN prediction.  Before this 
can be claimed as the resolution of the conflict, more data on stars 
in other Globular Clusters is required.  For a discussion of the 
proposed astrophysical solutions to this problem and further 
references, see\cite{steigman07}. 

\subsection{BBN Summary}

For SBBN, the only adjustable parameter is the baryon density parameter.
Using deuterium, the baryometer of choice, to fix this parameter it is
found (\S2.1) that
\be
10^{10}\eta_{\rm B}({\rm SBBN}) = \eta_{10}({\rm SBBN}) = 5.96^{+0.30}_{-0.33} \ ;
\ \ \Omega_{\rm B}h^{2}({\rm SBBN}) = 0.0218^{+0.0011}_{-0.0012}.
\ee
This value of $\eta_{10}$ is then used to find the BBN-predicted 
abundances of the other light nuclides, \3he, \4he, and \7li.  When 
the predictions are compared to the observations, there is excellent 
agreement for \3he (within very large uncertainties).  The predicted 
and observationally-inferred primordial abundances of \4he agree at 
a level of $\sim$1.4$\sigma$.  Lithium, however, poses a challenge;
the predicted abundance exceeds the observationally suggested
value by a factor of three or more.  We'll return to this problem
below, after discussing the effect of a non-standard expansion
rate on BBN and the CMB.

\section{CMB Constraint On the Baryon Density Parameter}

The relative heights of the odd and even ``acoustic" peaks in the
CMB temperature anisotropy spectrum are sensitive to the baryon
density parameter.  The CMB is a baryometer for the $\sim$400
thousand year old Universe.  The CMB determination of $\eta_{\rm B}$
is largely uncorrelated with $S$ (or, N$_{\nu}$), so that the value
determined for \Nnu = 3 is consistent with that for N$_{\nu} \neq 3$.
In our recent analysis, V.~Simha and I, combining the WMAP 5-year
data\cite{wmap} with other CMB experiments\cite{cmb}, supplemented by 
LSS\cite{lss} and SNIa\cite{snia} data, and the HST determination of the 
Hubble parameter\cite{hst}, needed to break cosmological degeneracies, 
found\cite{vs1},
\be
10^{10}\eta_{\rm B}({\rm CMB}) = \eta_{10}({\rm CMB}) = 6.14^{+0.16}_{-0.11} \ ; 
\ \ \Omega_{\rm B}h^{2}({\rm CMB}) = 0.0224^{+0.0006}_{-0.0004}.
\ee
The CMB/LSS data determine the baryon density parameter (at $\sim$400 
thousand years) to an accuracy of $\la 3\%$.  The excellent agreement 
between the BBN- and CMB-inferred values of the baryon density parameter 
is displayed in Figure \ref{fig:eta10sbbn}, where the solid curve shows the 
CMB-determined likelihood distribution (for further details and references, 
see\cite{vs1}).  Below, the implications of this excellent agreement for some 
classes of models of non-standard physics is discussed.  First, though, let's 
explore the effect on BBN and the CMB of a non-standard, early-Universe 
expansion rate.

\section{Non-Standard Particle Content (N$_{\nu} \neq 3$) Or Expansion 
Rate ($S \neq 1$)?}

\begin{figure}
\centerline{
\epsfxsize=3.6truein %4.2truein
\epsfbox{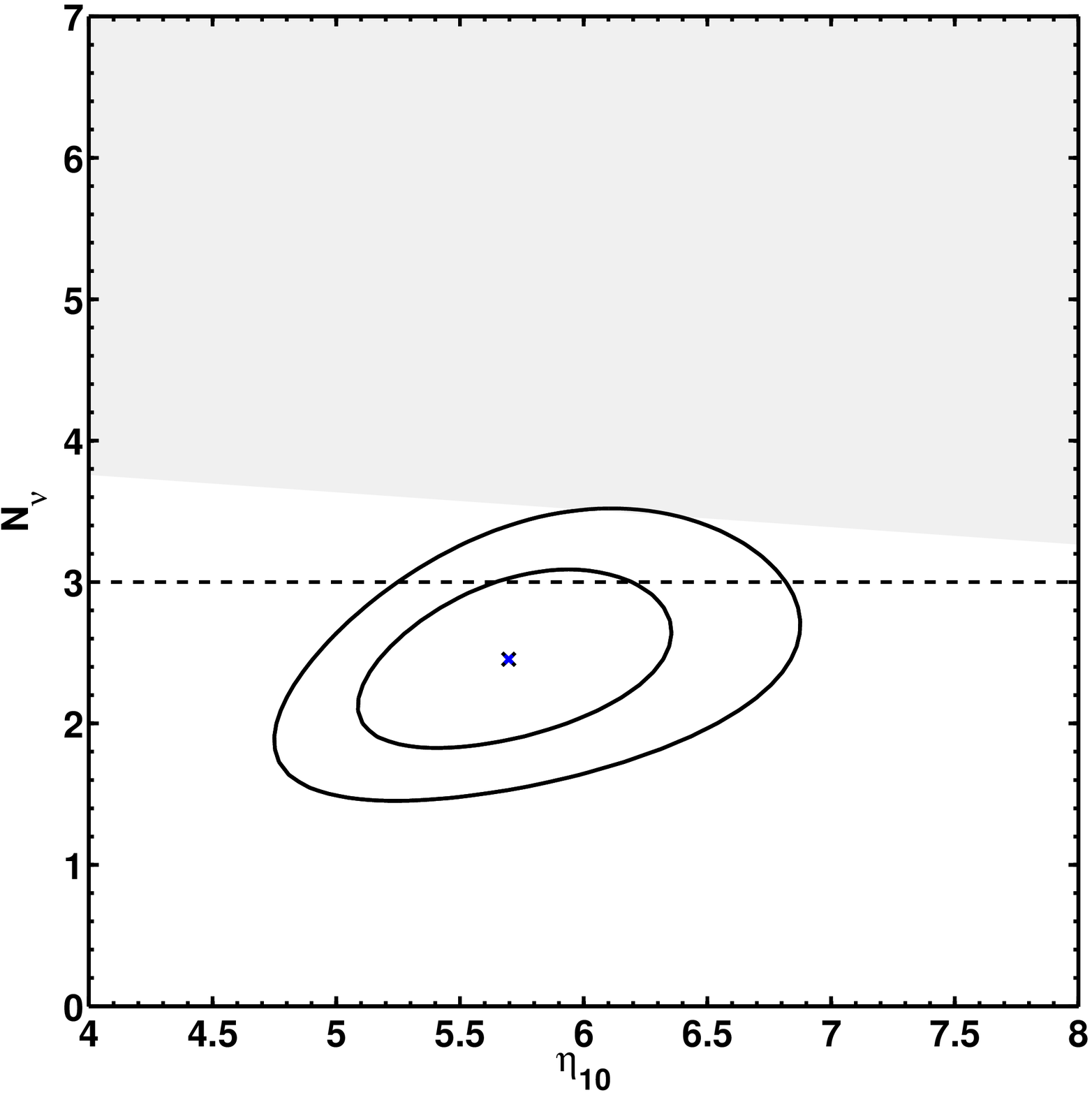} %[45 482 335 725]
}
\caption{}
The 68\% and 95\% contours in the \Nnu -- $\eta_{10}$ plane inferred 
from BBN using the D and \4he abundances adopted in \S2.1 and \S2.3
respectively.  The ``x" marks the ``best fit" point\cite{vs1} at $\eta_{10} 
= 5.7$ and \Nnu = 2.4.  The shaded area corresponds to the combination 
of \Nnu and $\eta_{10}$ leading to a BBN-predicted \4he mass fraction 
in excess of the observationally-inferred upper bound of 0.255; 
see \S2.3.
\label{fig:nnue10bbn}
\end{figure}

If the assumption of the standard model expansion rate is relaxed, both 
BBN and the CMB are affected.  For BBN, the largest effect is on the relic 
abundance of \4he, due to its sensitivity to the ratio of neutrons to protons 
at BBN, since virtually all neutrons available are incorporated into \4he, 
so that Y$_{\rm P} \approx [2n/(n+p)]_{\rm BBN}$.  In the early Universe, 
prior to BBN, when the temperature drops below $\sim$0.8 MeV, the rate 
of the charged-current weak interactions regulating the n/p ratio becomes 
smaller than the universal expansion rate, the Hubble parameter $H$.  
Thereafter, the n/p ratio deviates from (exceeds) its equilibrium value 
of $n/p =~$exp$(-\Delta m/kT)$, where $\Delta m$ is the neutron-proton mass 
difference.  The value of n/p when BBN begins is therefore determined by
the competition between the weak interaction rate and $H$.  An increase 
in the expansion rate ($S \geq 1$; $\Delta$N$_{\nu} \geq 0$) leaves less 
time for neutrons to convert to protons, increasing the predicted \4he 
primordial abundance.  According to Kneller \& Steigman\cite{ks}, a very 
good approximation to \Yp in this case, updated to account for incomplete 
neutrino decoupling\cite{mangano}, is\cite{steigman07,vs1}
\be
{\rm Y}_{\rm P} = 0.2485 \pm 0.0006 + 0.0016[(\eta_{10} - 6) + 100(S - 1)].
\label{y_p}
\ee
In contrast, the BBN-predicted abundances of the other light elements
are less sensitive to $S$.  For example, for deuterium\cite{ks,steigman07},
\be
y_{\rm D} = 2.64(1 \pm 0.03)\left[{6 \over \eta_{10} - 6(S - 1)}\right]^{1.6}.
\ee
For the ranges in $\eta_{10}$ and $S$ where these fits are accurate 
(within the quoted errors), $4~\la \eta_{10}~\la 8$ and $0.8~\la S~\la 
1.1$ ($0.8~\la $N$_{\nu}~\la 4.3$), the $y_{\rm DP}$ and \Yp isoabundance 
contours in the $S - \eta_{10}$ plane are nearly orthogonal, so that 
the observationally-inferred primordial D and \4he abundances serve to 
determine the $\{S,\eta_{10}\}$ pair.  In Figure \ref{fig:nnue10bbn} are 
shown the 68\% and 95\% contours in the \Nnu -- $\eta_{10}$ plane inferred 
from BBN and the D and \4he abundances adopted above.

For BBN constrained by D and \4he, Simha and Steigman\cite{vs1} find,
\be
\eta_{10}({\rm BBN})=5.7\pm0.4 \ , \ \ {\rm N}_{\nu}({\rm BBN})=2.4\pm0.4.
\ee
Although the ``best fit" value for \Nnu is less than the standard model 
value of \Nnu = 3, BBN and the D and \4he abundances are in agreement 
with the standard model at the $\sim$68\% confidence level.  While 
this leaves the excellent agreement with \3he unchanged, it fails to
resolve the ``lithium problem"\cite{vs1}. 

The CMB temperature anisotropy power spectrum is affected by the radiation 
density primarily through the effect of $\rho_{\rm R}$ in determining 
the epoch (redshift) of matter-radiation equality.  The amplitudes of 
fluctuations on scales which enter the horizon when the Universe is 
still radiation dominated differ from the amplitudes on those scales 
which enter the horizon later, when the Universe is matter dominated.  
Increasing the radiation content (\Nnu $\geq 3$) delays matter-radiation 
equality, leaving less time before recombination, suppressing the growth 
of perturbations.  The redshift of the epoch of matter-radiation equality, 
$z_{eq}$, is related to the matter (baryonic plus non-baryonic) and 
radiation densities by,
\begin{equation}
1+z_{eq} = \rho_{\rm M}/\rho_{\rm R}.
\end{equation}
Since $\rho_{\rm R}$ depends on N$_{\nu}$, $z_{eq}$ is a function of both 
\Nnu and $\Omega_{\rm M}h^2$, leading to a degeneracy between these two 
parameters.  The CMB constrains $z_{eq}$, so that any increase in \Nnu 
needs to be compensated by a corresponding increase in $\Omega_{\rm M}h^2$.  
As a result of this degeneracy, the CMB power spectrum alone can only lead 
to a very weak constraint on N$_{\nu}$.  Constraints on these parameters 
independent of the CMB are needed to break the degeneracy between them. 

Since the luminosity distances of type Ia supernovae (SNIa)\cite{snia} provide 
a constraint on a combination of $\Omega_{\rm M}$ and $\Omega_{\Lambda}$ 
complementary to that from the assumption of flatness, they are of value 
in restricting the allowed values of $\Omega_{\rm M}$.  In concert with 
a bound on $H_{0}$\cite{hst}, this helps to break the degeneracy between \Nnu 
and $\Omega_{\rm M}h^{2}$. 
 
Another way to break the degeneracy between \Nnu and $\Omega_{\rm M}h^2$ 
is to use LSS measurements of the matter power spectrum in combination 
with the CMB power spectrum.  The turnover scale in the matter power 
spectrum is set by the requirement that $z_{eq}$ remain unchanged as
\Nnu deviates from its standard-model value of 3.  Since the baryon 
density is constrained by the CMB power spectrum, independently of 
N$_{\nu}$, increasing the radiation density (\Nnu $> 3$) requires a 
higher dark matter density in order to preserve $z_{eq}$ (in a flat 
universe, $\Omega_{\rm M} + \Omega_{\Lambda} = 1$).  Between the epoch 
of matter-radiation equality and recombination the density contrast 
in the cold dark matter grows unimpeded, while the baryon density 
contrast cannot grow.  Consequently, increasing \Nnu and $\Omega_{\rm M}h^2$ 
increases the amplitude of the matter power spectrum on scales smaller 
than the turnover scale corresponding to the size of the horizon at 
$z_{eq}$.  In this way, data from galaxy redshift surveys\cite{lss} can 
be used to infer the matter power spectrum, thereby constraining 
$\Omega_{\rm M}h^2$ and N$_{\nu}$.
\begin{figure}
\centerline{
\epsfxsize=2.95truein
\epsfbox{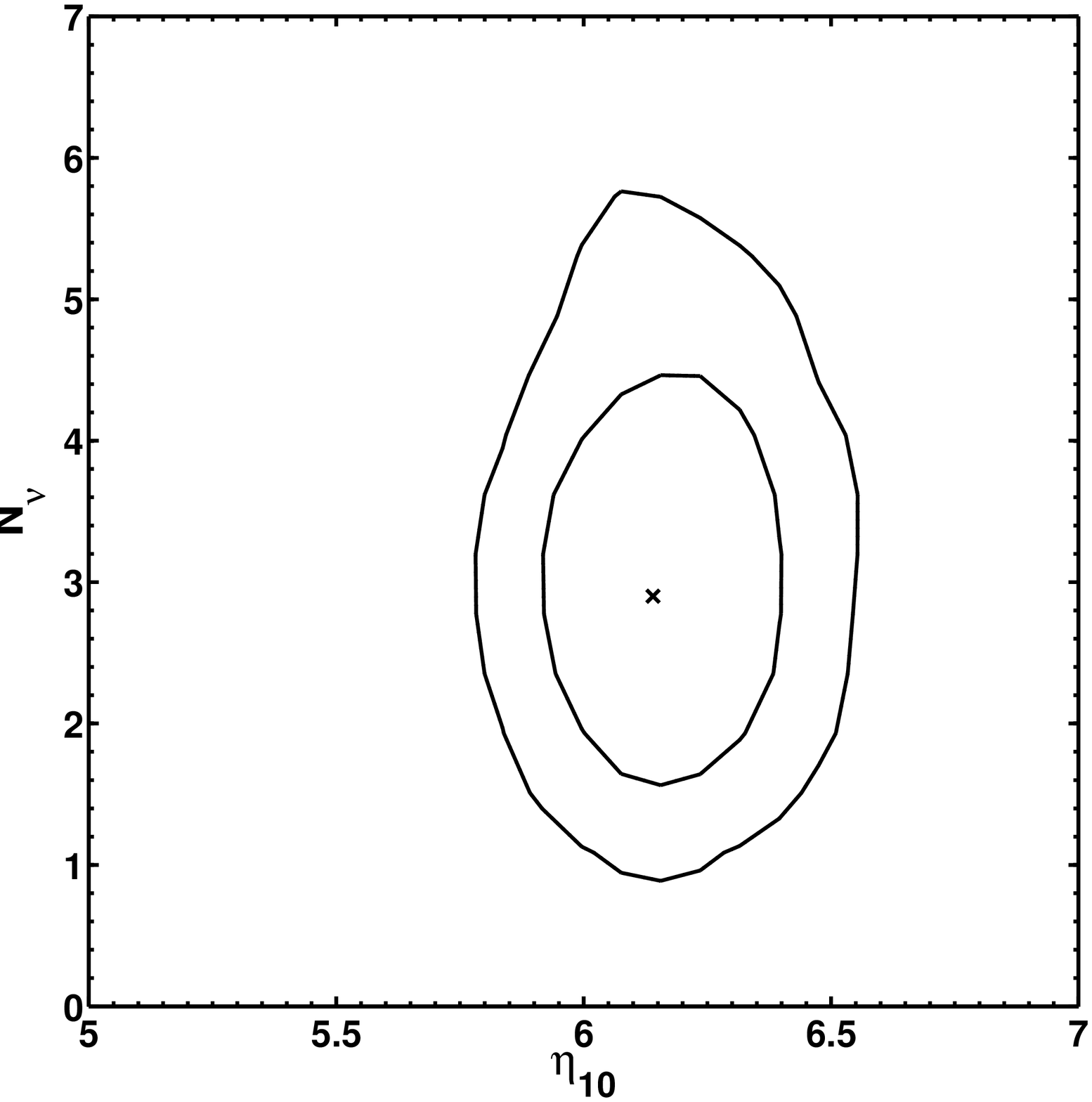}
}
\vspace*{13pt}
\caption{}
The 68\% and 95\% contours in the N$_{\nu} - \eta_{10}$ plane derived 
using the WMAP 5-year data, small scale CMB and SNIa data, and the 
HST Key Project prior on $H_0$ along with data from the LSS matter 
power spectrum (see the text).
\label{fig:nnue10cmblss}
\end{figure}
For these complementary datasets, Simha \& Steigman\cite{vs1} obtain,
\be
\eta_{10}({\rm CMB})=6.1^{+0.2 +0.3}_{-0.1 -0.2},
\ee
and
\be
{\rm N}_{\nu}({\rm CMB})=2.9^{+1.0 +2.0}_{-0.8 -1.4}.
\ee
The 68\% and 95\% contours in the \Nnu -- $\eta_{10}$ plane are shown
in Figure \ref{fig:nnue10cmblss}, where it is clear that, so far, the 
CMB is a better baryometer than a chronometer.

\section{Comparing BBN With The CMB}

\begin{figure}
\centerline{\epsfxsize=5.25truein\epsffile{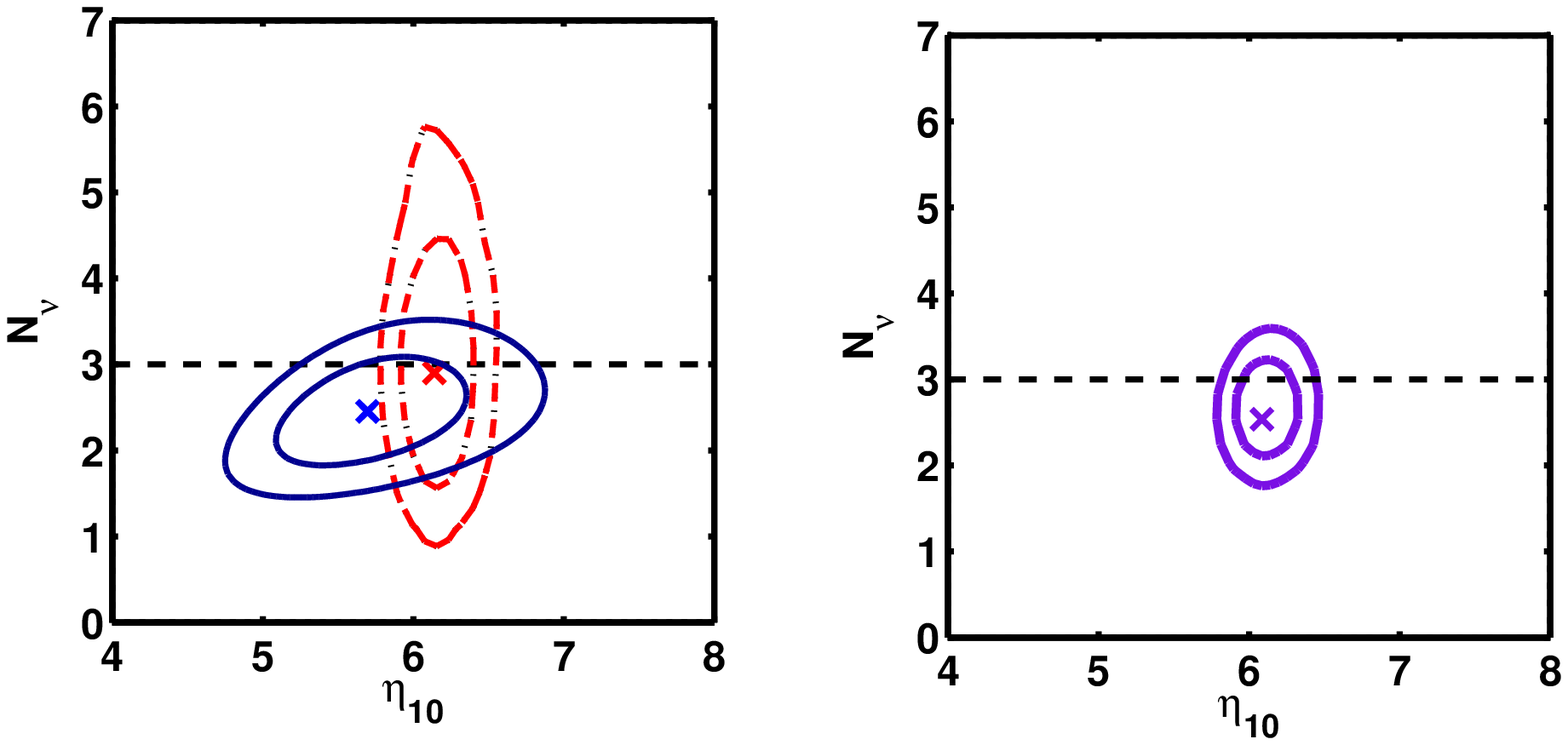}}
\vspace*{13pt}
\caption{}
(Left) In blue (solid), the 68\% and 95\% contours in the \Nnu - $\eta_{10}$ 
plane derived from a comparison of the observationally-inferred and 
BBN-predicted primordial abundances of D and \4he.  In red (dashed), 
the 68\% and 95\% contours derived from the combined WMAP 5-year data, 
small scale CMB data, SNIa, and the HST Key Project prior on $H_0$ along 
with the LSS matter power spectrum data. (Right) The 68\% and 95\% joint 
BBN-CMB-LSS contours in the N$_{\nu} - \eta_{10}$ plane.
\label{fig:joint}
\end{figure}

At present the combined CMB and LSS data provide the best baryometer,
determining the baryon density to better than 3\%, but only a relatively 
weak chronometer, still allowing a large range in $S$ ($0.87 \leq S \leq 
1.14$ or, $1.5 \leq$ N$_{\nu} \leq 4.9$ at 95\% confidence).  BBN (D \& 
\4he) provides a consistent, but tighter constraint on $S$ ($0.88 \leq 
S \leq 1.02$ or, $1.6 \leq$ N$_{\nu} \leq 3.2$ at 95\% confidence).  As 
may be seen from the left hand panel of Figure~\ref{fig:joint}, within 
the uncertainties, the CMB/LSS, which probes the Universe at $\ga 400$ 
thousand years, is consistent with BBN, which provides a glimpse of the 
Universe at $\la 20$ minutes.  For example, the BBN abundances of D, \3he, 
and \4he, inferred using the CMB/LSS values of $\eta_{10}$ and N$_{\nu}$, are 
in excellent agreement, within the errors, with the observationally-inferred 
relic abundances.  Lithium, however, remains a problem.   
The excellent agreement between BBN and the CMB permits constraints on 
any deviations from standard-model physics between BBN and recombination 
and/or the present epoch.  For example, since baryons are conserved, 
$\eta_{\rm B}$ relates the number of thermalized black body photons in a 
comoving volume at different epochs, constraining any post-BBN entropy 
production.  The consistency of $\eta_{\rm B}$ inferred from the CMB/LSS 
and from BBN implies\cite{vs1},
\be
{N_{\gamma}^{\rm CMB} \over N_{\gamma}^{\rm BBN}} = 0.92 \pm 0.07.
\ee 
This ratio is consistent with no change in entropy between BBN and
recombination at $\sim$1$\sigma$, placing an interesting upper bound 
on any post-BBN entropy production.

For another example, late decaying particles might produce additional 
relativistic particles (radiation), but not thermalized, black body 
photons\cite{ichikawa07}.  Such deviations from the standard model 
radiation density can be parameterized by the ratio of the radiation 
density, $\rho'_{\rm R}$, to the standard model radiation density, 
$\rho_{\rm R}$ (or, to the photon energy density $\rho_{\gamma}$).  
In the post-\epm annihilation universe relevant for this comparison
of BBN with the CMB,
\be
R = S^2 = {\rho'_{\rm R} \over \rho_{\rm R}} = 1 + 0.134\Delta{\rm N}_{\nu}.
\ee
Comparing this ratio ($R$) at BBN and at recombination constrains 
any post-BBN production of relativistic particles.
\be
{R_{\rm CMB} \over R_{\rm BBN}} = 1.07^{+0.16}_{-0.13}
\ee
This ratio, too, is consistent with unity within 1$\sigma$, constraining 
any post-BBN relativistic particle production.

While a non-standard expansion rate ($S \neq 1$) has been parameterized 
in terms of an equivalent number of neutrinos (N$_{\nu} \neq 3$), we 
reiterate that a non-standard expansion rate need not be due to the 
presence of extra (or fewer) neutrinos.  For example, deviations from 
the standard expansion rate could occur if the value of the early-Universe 
gravitational constant, $G$, were different in the from its present, 
locally-measured value $G_{0}$.  For the standard radiation density 
with three species of light, active neutrinos, the constraint on the 
expansion rate can be used to constrain variations in the gravitational 
constant, $G$.  From BBN, 
\be
G^{\rm BBN}/G_{0} = S^{2}_{\rm BBN} = 0.91 \pm 0.07
\ee 
and at the epoch of the recombination,
\be
G^{\rm CMB}/G_{0} = S^{2}_{\rm CMB} = 0.99^{+0.13}_{-0.11}.
\ee
Both are consistent with no variation in G at the $\sim$1$\sigma$ level.

Since the independent constraints on $\eta_{\rm B}$ and \Nnu from BBN and
the CMB/LSS are in very good agreement, they may be combined to obtain
the joint fit shown in the right hand panel of Figure \ref{fig:joint}.  
By comparing the two panels of Figure \ref{fig:joint}, it may be seen
that the CMB/LSS drives the constraint on the baryon density parameter,
while BBN is largely responsible for the bounds on N$_{\nu}$.   At 68\% 
and 95\% confidence, Simha and Steigman\cite{vs1} find
\be
\eta_{10}=6.11^{+0.12 +0.26}_{-0.13 -0.27} \ ; \ \ 
\Omega_{\rm B}h^{2} = 0.0223^{+0.0004 +0.0009}_{-0.0005 -0.0010},
\ee
and
\be
{\rm N}_{\nu}=2.5\pm0.4\pm0.7.
\ee
While the best-fit value of \Nnu is less than $3$, this difference 
is not statistically significant.

\section{Conclusions}

In the standard models of cosmology and of particle physics the
particle content is fixed (\eg \Nnu = 3) and, baryon number has been 
conserved since the very earliest epochs.  By comparing BBN with 
the CMB and LSS, baryon conservation can be tested between $\sim$20
minutes (BBN) and $\sim$400 thousand years.  So, too, can deviations
from the standard-model particle content (\Nnu $\neq 3$?), as well as
deviations from the standard-model predicted early Universe expansion
rate.  In this talk and review, $\eta_{\rm B}$ evaluated from BBN and
from the CMB have been compared and shown to be in excellent agreement,
consistent with the standard model expectation.  BBN and the CMB have
also been employed to compare \Nnu (or, $S$) at these two, widely
separated epochs and, to compare them to the standard-model expectation
that \Nnu = 3 ($S=1$).  While the central values of \Nnu evaluated at BBN 
and from the CMB differ from each other and from the standard-model 
value, they are, in fact, in agreement with each other, and with the
standard-model value within the uncertainties.  This concordance of the 
standard models of particle physics and cosmology permit constraints on 
``new" physics at the times (and energies) between BBN and the CMB (and, 
at present).

Prior to WMAP and the other ground- and balloon-based CMB experiments, 
BBN and deuterium provided the best cosmological baryometer, while \4he 
provided the only early-Universe chronometer.  The WMAP 5-year data, 
combined with other CMB and LSS data, now lead to a determination of 
the baryon density at the $\sim$2-3\% level, a factor of $\sim$2 better 
than that from BBN.  However, although the CMB/LSS constraint on \Nnu has 
improved significantly and is consistent with that from BBN, it still 
remains weaker than the corresponding BBN constraint.  For some time now 
BBN has established at high confidence that \Nnu $> 1$ at BBN, when the 
Universe was $\sim$20 minutes old.  The more recent CMB data, combined 
with other SNIa and LSS datasets, now confirm that \Nnu $> 1$ (or, \Nnu $> 
2$~\cite{dunkley08}) when the Universe was $\ga 400$~kyr old.

It may be of interest to some readers that, since the NO-VE IV Workshop, 
V. Simha and I have presented the results of a similar analysis, updating 
the constraints from BBN and the CMB on the lepton asymmetry parameter\cite{vs2}.
  
\section{Acknowledgements}
  Words alone cannot express my admiration for and gratitude to this 
  workshop's organizer Milla Baldo Ceolin and her colleagues.  I thank 
  Vimal Simha for discussions and for preparing the figures used here. 
  The research summarized here is supported at The Ohio State University 
  by a grant from the US Department of Energy.

\end{document}